\documentclass[10pt]{article}
\usepackage{style} 
\usepackage{url}
\usepackage{latexsym}
\usepackage{amsmath, amsthm, amsfonts}
\usepackage{algorithm, algorithmic}  
\usepackage{graphicx}
\usepackage{lipsum} 
\usepackage{fancyhdr}
\usepackage[headheight=1in,margin=1in]{geometry}
\usepackage{hyperref}
\usepackage[T1]{fontenc}
\usepackage{enumitem}
\title{Science Fiction and Fantasy in Wikipedia: Exploring Structural and Semantic Cues}

\author{
  Włodzimierz Lewoniewski  \\
  Poznań University \\of Economics \\ and Business \And  
  Milena Stróżyna \\
  Poznań University \\of Economics \\ and Business \And
  Izabela Czumałowska \\
  Poznań University \\of Economics \\ and Business \And 
  Elżbieta Lewańska \\
  Poznań University \\of Economics \\ and Business  }

\begin{document}
\maketitle
\thispagestyle{fancy}

\begin{abstract}
Identifying which Wikipedia articles are related to science fiction, fantasy, or their hybrids is challenging because genre boundaries are porous and frequently overlap. Wikipedia nonetheless offers machine-readable structure beyond text, including categories, internal links (wikilinks), and statements if corresponding Wikidata items. However, each of these signals reflects community conventions and can be biased or incomplete. This study examines structural and semantic features of Wikipedia articles that can be used to identify content related to science fiction and fantasy (SF/F).
\end{abstract}

\section*{Introduction}
Classifying Wikipedia articles as science fiction, fantasy, or mixed forms is more challenging than it initially appears. In genre studies, the boundary between science fiction and fantasy is widely treated as porous, historically contingent, and rhetorically negotiated rather than fixed by necessary-and-sufficient conditions. For example, scholarship on speculative narratives emphasizes that science fiction and fantasy show no single stylistic uniformity, can coexist within the same work, and often overlap with adjacent modes such as mythic, utopian/dystopian, horror, or “science fantasy”. This conceptual ambiguity creates a practical labeling problem on Wikipedia: many pages are clearly in-scope (e.g., well-known novels, films, authors), while others occupy gray zones (e.g., folklore retellings with technoscientific framing; magical realism with speculative elements; role-playing game settings; fandom and production artifacts).

At the same time, Wikipedia is an attractive corpus for large-scale genre mapping because it provides rich, machine-readable structure beyond plain text: category assignments, in-article hyperlinks, and cross-project metadata such as Wikidata alignment and WikiProject tags. However, each signal encodes editor practices and community conventions, which introduces bias, incompleteness, and incompatibilities that complicate any attempt to build a unified, reproducible definition of Wikipedia articles related to “Science Fiction and Fantasy” (SF/F).

The goal of this research is to identify structural and semantic features of Wikipedia that can be used to detect articles related to science fiction and fantasy. Concretely, we examine which signals are consistently characterize SF/F-relevant articles drawn from Wikipedia’s category structure, wikilinks from lead section, Wikidata alignment. Extraction of wikilinks, article categories, Wikidata statements was performed offline using public Wikimedia Dumps as of 1 January 2026.\footnote{\url{https://dumps.wikimedia.org/}}

\section*{WikiProjects as a Baseline}
We began from analysis of WikiProject membership as a community-curated approximation of topical relevance. Using three science fiction and fantasy related WikiProjects as seeds, we obtained the following subsets\footnote{We selected only pages from the main namespace (NS=0). Redirects were excluded; instead, we included their target pages.}:
\begin{itemize}[noitemsep]
	\item \textbf{Fantasy}: WikiProject Novels/Fantasy task force\footnote{\url{https://en.wikipedia.org/wiki/Wikipedia:WikiProject_Novels/Fantasy_task_force}} - 4,355 articles,
	\item \textbf{Science Fiction}: WikiProject Science Fiction\footnote{\url{https://en.wikipedia.org/wiki/Wikipedia:WikiProject_Science_Fiction}} - 11,930 articles,
	\item \textbf{Science Fiction Novels}: WikiProject Novels/Science fiction task force\footnote{\url{https://en.wikipedia.org/wiki/Wikipedia:WikiProject_Novels/Science_fiction_task_force}} - 4,617 articles.
\end{itemize}

WikiProject tags are placed usually in talk pages of the Wikipedia articles and are not mutually exclusive: a single article can be tagged by multiple projects, which is useful for representing genre hybridity but complicates downstream evaluation. Therefore, total unique titles in three previously described subsets is 18,829 (instead of 20,902) - this is SF/F baseline set.

\section*{Wikidata}
\begin{figure*}[ht]
	\begin{center}
		\centerline{\includegraphics[width=\textwidth]{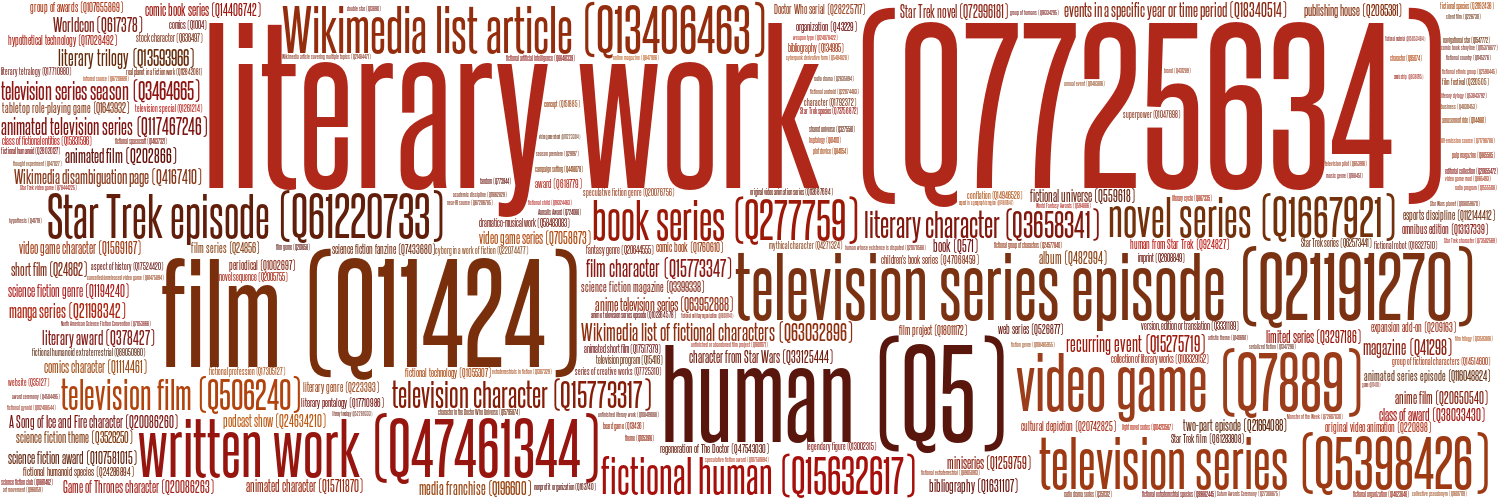}}
		\caption{The most frequent values in Wikidata items related to "Science Fiction and Fantasy" (SF/F) baseline set.}
		\label{fig:wikidata}
	\end{center}
\end{figure*}

We aligned each article to its corresponding Wikidata item and extract the set of “instance of” statements (P31) for each of them. This statement provides explicit typing (e.g., novel, film, fictional character, TV series, mythological creature). The figure \ref{fig:wikidata} shows most frequent values in P31 property in Wikidata items related to SF/F baseline set. For instance, the most popular type of described objects is “literary work” (Q7725634) - 38.54\% of articles in the SF/F baseline set and the highest share in Fantasy subset (65.56\%). Another popular type is “film” (Q11424), with 18.40\% share in Science Fiction subset.

\section*{Wikipedia categories}
\begin{figure*}[ht]
	\begin{center}
		\centerline{\includegraphics[width=\textwidth]{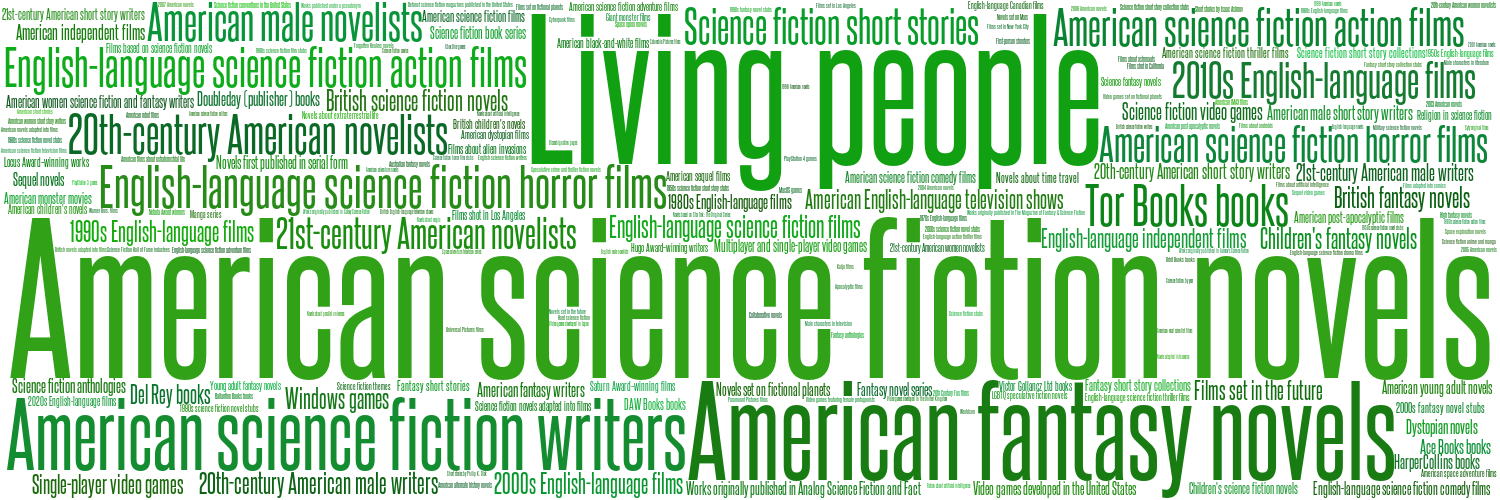}}
		\caption{The most frequent categories assigned to Wikipedia articles from SF/F baseline set.}
		\label{fig:categories}
	\end{center}
\end{figure*}

We collected the set of categories each article belongs to and analyze their distributions within our baseline corpora. The figure \ref{fig:categories} shows most frequent categories assigned to Wikipedia articles from SF/F baseline set.

It is important to note that, even when categories have explicitly science-fiction or fantasy–related names, not all articles in those categories are tagged by the corresponding WikiProjects. For instance, the following statistics summarize coverage for several high-popularity categories:
\begin{itemize}[noitemsep, topsep=0pt]
\item \textbf{American science fiction novels}: 1,257 articles are in the SF/F set (out of 1,918 total).
\item \textbf{American fantasy novels}: 1,026 articles are in the SF/F set (out of 1,532 of total).
\item \textbf{American science fiction writers}: 603 articles are in the SF/F set (out of 1,536 total).
\end{itemize}

\section*{Wikilinks}
\begin{figure*}[ht]
	\begin{center}
		\centerline{\includegraphics[width=\textwidth]{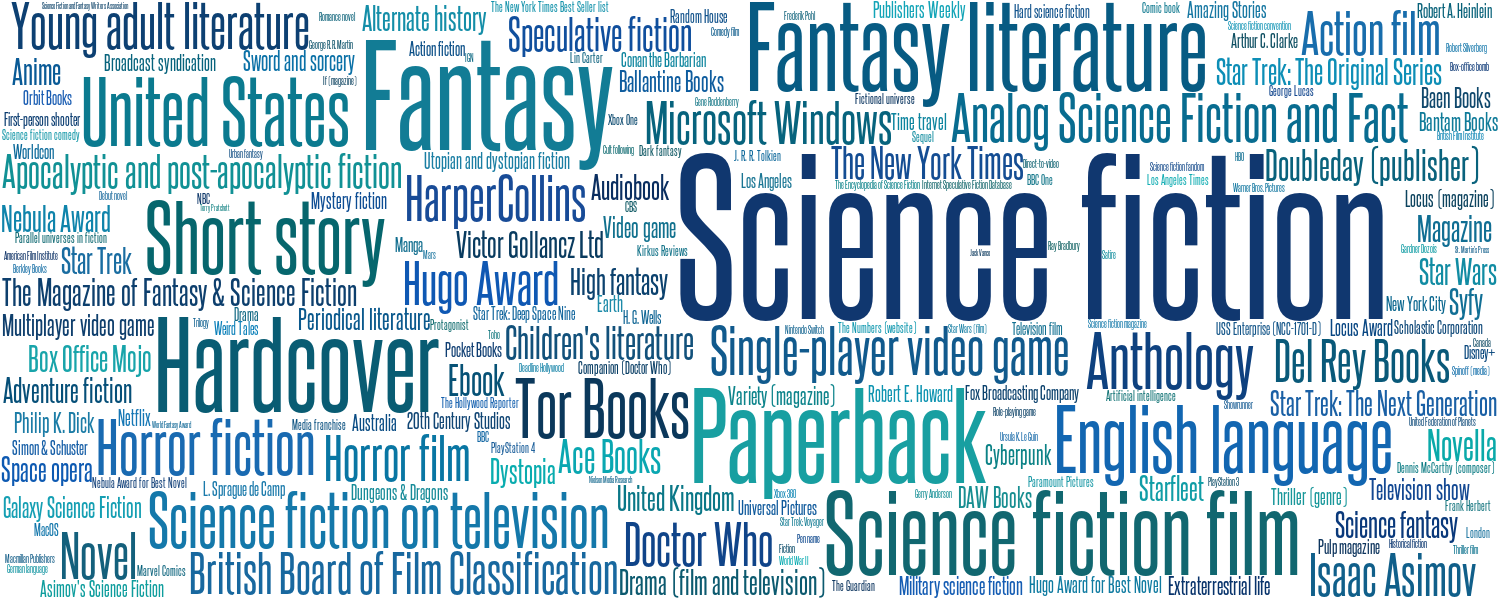}}
		\caption{The most frequent wikilinks in lead section of Wikipedia articles from SF/F baseline set.}
		\label{fig:wikilinks}
	\end{center}
\end{figure*}

Finally, we extract internal wikilinks appearing in the lead section, including infobox(es) in it, using source of the Wikipedia articles in wikimarkup. This choice is motivated by the editorial function of these regions: they are designed to summarize and contextualize the topic quickly, and empirical work on Wikipedia navigation shows that lead/infobox links tend to point toward more general, definitional concepts than links in later sections. The figure \ref{fig:wikilinks} shows most frequent wikilinks in lead section of Wikipedia articles from SF/F baseline set.

Let’s consider a few examples. In our SF/F set 7,066 articles link to the Wikipedia article “Science fiction” from their lead section, whereas across the entire English Wikipedia the number of articles linking to “Science fiction” from the lead section is 14,405 (i.e., our baseline covers about 49\% of such lead-linking pages). Similarly, 2,377 articles in our SF/F set link to “Fantasy” from the lead section, compared to 8,510 lead-linking articles in English Wikipedia overall (about 28\% coverage).

\section*{Discussion and Limitations}
During our research we found that identifying Wikipedia articles related to science fiction and fantasy cannot be reduced to a single reliable cue. Even with WikiProject-derived seed sets, “genre-relatedness” is often expressed indirectly, and different Wikipedia signals capture different facets of the same topic—sometimes agreeing on core pages, but diverging on boundary cases. WikiProject coverage is uneven and sometimes organized around franchises rather than genres. Well-known series can be curated under dedicated projects without consistent tagging by broader SF/F projects, which can leave gaps if only general WikiProjects are used. For example, “Game of Thrones” is aligned with WikiProject “A Song of Ice and Fire” and is not aligned with any of the three WikiProjects considered in this study.

Wikidata adds structured semantics (e.g., P31 \textit{instance of}), but statement completeness and modeling choices vary: items may have multiple P31 values, and typing can be inconsistent across domains. Consequently, Wikidata signals can be informative yet noisy, especially for complex entities like franchises, universes, and adaptations.

Categories provide explicit editorial grouping, but they are not a clean taxonomy and SF/F relevance may appear only through subcategories or parent categories. Shallow category use can miss relevant pages, while deeper traversal can pull in peripheral topics (publishing, awards, fandom), so results depend strongly on traversal depth and pruning rules.

Wikilinks from leads and infoboxes tend to emphasize broad context (e.g., countries, awards, publishers, occupations) because these sections are written for quick orientation. This makes them useful for coarse “aboutness”, but can be less reliable for distinguishing genre, as niche subgenres may be underlinked while highly connected non-genre entities are overrepresented.

\section*{Future Work}
In future works we plan to extend articles features, including quality and popularity measures \cite{lewoniewski2019,lewoniewski2025}, as well as consider other statements from corresponding Wikidata items (e.g., P136, \textit{genre}). We plan to add more language version to the analysis. More extended results of this research can be found at: \url{https://data.lewoniewski.info/fantasy}

\bibliographystyle{style} 
\bibliography{references}

\clearpage
\newpage

\end{document}